\documentclass[prb,twocolumn,showpacs,preprintnumbers,amsmath,amssymb]{revtex4}

\usepackage{graphicx}
\usepackage{dcolumn}
\usepackage{bm}%
\usepackage{amsmath}%
\usepackage{amsfonts}%
\usepackage{amssymb}
\newcommand{\lgun}{\lambda_{\rm gun}}

\newcommand{\Bz}{B_{\rm Z}}

\newcommand{\Bt}{B_\phi}

\begin{document}
\title{On the jets, kinks, and spheromaks formed by a planar
magnetized coaxial gun}
\author{S.~C.~Hsu}
\thanks{present address: Los Alamos National Laboratory, Los Alamos, NM\, 87545}
\email{scotthsu@lanl.gov}
\author{P.~M.~Bellan}
\email{pbellan@caltech.edu}
\affiliation{California Institute of Technology, Pasadena, CA\, 91125}

\date{\today}

\begin{abstract}
Measurements of the various plasma configurations produced by a planar
magnetized coaxial gun provide insight into the magnetic topology evolution
resulting from magnetic helicity injection. Important features of the
experiments are a very simple coaxial gun design so that all observed
geometrical complexity is due to the intrinsic physical dynamics rather than
the source shape and use of a fast multiple-frame digital camera which
provides direct imaging of topologically complex shapes and dynamics. Three
key experimental findings were obtained: (1)~formation of an axial collimated
jet [Hsu and Bellan, Mon.\ Not. R.\ Astron.\ Soc.~\textbf{334}, 257 (2002)]
that is consistent with a magnetohydrodynamic description of
astrophysical jets,
(2)~identification of the kink instability when this jet satisfies the
Kruskal-Shafranov limit, and (3)~the nonlinear properties of the kink
instability providing a conversion of toroidal to poloidal flux as required
for spheromak formation by a coaxial magnetized source [Hsu and Bellan,
Phys.\ Rev.\ Lett.~\textbf{90}, 215002 (2003)].  A new interpretation
is proposed for how the $n=1$ central column instability
provides flux amplification during spheromak formation and sustainment,
and it is shown that jet collimation can occur within one rotation
of the background poloidal field.

\end{abstract}
\pacs{52.55.Ip,52.72.+v,52.35.Py,95.30.Qd}
\maketitle


\section{\label{sec:intro} Introduction}

There has long been a mutually beneficial exchange of ideas between laboratory
plasma physics and plasma astrophysics. An example is the proposal by
Lundquist\cite{lundquist50} that cosmical magnetic fields are
force-free,\cite{chandrasekhar58} satisfying $\nabla\times\mathbf{B} =
\lambda\mathbf{B}$. Woltjer showed\cite{woltjer58a} that these force-free
states are intimately related to the concept of magnetic helicity conservation
and that the force-free state results from a variational principle in which
the magnetic energy of a system is minimized subject to the constraint of
constant magnetic helicity, which is a measure of magnetic flux twist and
linkage.\cite{Berger84} This idea was then used to explain the spontaneous
generation of reversed magnetic fields in the reversed field
pinch\cite{Taylor74} and the tendency of coaxial gun produced plasmas to reach
the spheromak state.\cite{turner83} Although \emph{Taylor relaxation}%
\cite{taylor86} as this process is now known, is not a complete and rigorous
theory, it was very successful, indicating that the same fundamental physical
principles govern both laboratory and astrophysical plasmas. The Taylor
argument showed that the end result of complex dynamics is a relatively simple
{\em relaxed} state, but the argument sidesteps the issue of characterizing
the detailed dynamical processes leading to this state. This paper describes
recent experiments providing new insights into the detailed dynamical processes
of magnetic helicity injection and plasma relaxation.  In particular,
the experiments shed light on the non-axisymmetric
dynamics of spheromak formation, a subject with
a long history beginning with the work of Alfv\'{e}n\cite{alfven60} and
Lindberg.\cite{lindberg61}  These experiments also
show how magnetic helicity injection can lead to the formation and
collimation of
magnetically-driven jets in astrophysics.\cite{begelman84,blandford00,meier01}

An important feature of this experiment is the use of the simplest possible
setup for allowing studies of the details and three-dimensional (3D) dynamics
of magnetic helicity injection and spheromak formation. Key results
include (1)~observation of distinct
plasma regimes depending on the parameter $\lambda_{\mathrm{gun}}$ (ratio of
peak gun current to flux), similar to those previously found\cite{yee00} using
a conventional cylindrical coaxial gun, (2)~formation and axial
expansion of a central plasma column from
merging of plasma-filled poloidal flux tubes\cite{hsu02b}
in a manner consistent with magnetohydrodynamic (MHD)
models of magnetically-driven astrophysical jets,\cite{lynden-bell96,meier01,
ouyed03,bellan03} (3)~clear identification of 
a kink instability of this central column and a kink onset threshold that is
quantitatively consistent with the Kruskal-Shafranov limit for an ideal MHD
kink,\cite{hsu02b,hsu03} and (4)~identification of the nonlinear limit of the
kink instability as a poloidal flux amplification mechanism leading to
spheromak formation.\cite{hsu03}

This paper is organized as follows.
Section~\ref{setup} describes the experimental setup. Section~\ref{results}
presents the experimental results. Section~\ref{discussion} provides
a detailed discussion of the results in the context of both spheromak
formation and astrophysical jets. A summary is given in Sec.~\ref{summary}.

\section{Experimental setup}

\label{setup}

In order to facilitate interpretation of results, the coaxial gun design was
made to be as geometrically simple as possible: a disk surrounded by a
co-planar annulus (Fig.~1) and a bias magnetic field
coil located just behind the disk (not visible in Fig.~1).
This experimental
setup provided the
additional advantage that the entire plasma formation region is in plain view
so that all steps of the topological evolution could be observed. 
Note that this setup has essentially the same magnetic topology
as a star-disk system associated with magnetically-driven
astrophysical jets (see Fig.~2).  The various
components of the experimental setup are described in detail below.

\subsection{Vacuum chamber}

The cylindrical vacuum chamber, shown in
Fig.~3(top), is about 2~m long, 1.4~m in diameter, and
constructed using 6~mm thick electro-polished stainless steel. The chamber has
large windows on the side that provide a good view of the entire plasma. A key
point is that the chamber size is large compared to both the coaxial gun
source and
the plasma, so that plasma-wall interactions are minimal during the time of
interest for the results presented here. The chamber is evacuated to $1.5
\times10^{-7}$~T with a cryogenic pump, and a sorption pump is used for initial
pump-down from atmosphere to 100~mT so that the the entire pumping system is
oil-free. The oil-free property and the minimal plasma-wall interactions
resulted in plasmas that did not require a discharge cleaning protocol in
order to provide clean, reproducible results.

\subsection{Planar magnetized coaxial gun}

The planar magnetized coaxial gun,\cite{hsu02a}
shown in Figs.~1
and 3(bottom), is mounted on one end-dome of the vacuum chamber, as
shown in Fig.~3(top). The gun consists of (1)~an inner 20.3~cm
diameter copper disk electrode, (2)~a co-planar, coaxial outer 50.8~cm
diameter copper annulus electrode, and (3)~a magnetic field coil located just
behind the inner electrode. The copper disk inner electrode is bolted onto
the vacuum-face of a re-entrant 7.5~in.\ diameter stainless steel half nipple
mounted on one end of an 8~in.\ diameter ceramic break which insulates the
inner electrode from ground. The other end of the ceramic break is mounted on
a 10~in.\ flange centered on the vacuum chamber end-dome. Thus, the inner
electrode is electrically floating and is connected to a capacitor bank via
coaxial cables connected across the ceramic insulating break. The outer
annulus electrode is mounted via copper brackets directly to the inner surface
of the vacuum chamber end-dome and therefore is electrically tied to chamber
ground. A 6~mm vacuum gap separates the disk and outer annulus. The
experiments are characterized using the cylindrical coordinate system shown in
Fig.~3(bottom); the $\phi$ and ($R$-$Z$) directions will be 
referred to as toroidal and poloidal, respectively.
The disk and annulus each have eight 6~mm
diameter gas injection holes spaced equally in $\phi$. The gas injection holes
on the disk and annulus are at $R=4.8$ and 17.8~cm, respectively.
Details of gas injection are given in
Sec.~\ref{gas}. The external bias magnetic field coil is located just behind
the central disk but, because of the re-entrant geometry, is at atmosphere.
The coil consists of 110 turns of \#11 insulated square magnetic wire wound on
a phenolic form (13.0~cm inner and 18.4~cm outer diameters)
and has an inductance of 2.3~mH
and resistance of 0.39~$\Omega$. Contours of constant poloidal flux $\psi$
generated by the coil are shown in Fig.~3(bottom).

\subsection{Power systems}

The coaxial gun is powered by an ignitron-switched  120~$\mu$F capacitor bank
rated up to 14~kV\@. Eight 2~m long Belden YK-198 low inductance coaxial
cables in parallel carry power from the capacitor bank to the gun electrodes.
Routine operation at bank voltages of 5~kV yields peak plasma currents
$I_{\mathrm{gun}} \approx80$~kA and gun voltages $V_{\mathrm{gun}} \approx
2.5$~kV (after breakdown), or peak gun power on the order of 200~MW\@. Typical
$I_{\mathrm{gun}}$ and $V_{\mathrm{gun}}$ traces are shown in
Fig.~4.

The external magnetic field coil is powered by a 1.4~mF, 450~V capacitor bank.
The half-period of the current trace for the external coil is 10~ms and is
effectively constant on the 10~$\mu$s time-scale of the experiment. The bias
poloidal magnetic flux $\psi_{\mathrm{gun}}$ (intercepting the inner gun
electrode) created by the external coil is typically 1--2~mWb.

\subsection{Gas injection}

\label{gas}

Fast gas puffs are required in these experiments because local gas pressures
on the order of 100~mT are required for breakdown. A slow or steady gas fill
is intolerable because the entire vacuum chamber would rise to 100~mT which
would result in a weakly ionized cold plasma and also an overloading of the
cryopump. Fast gas valves provide a transient, highly localized cloud of high
pressure gas in front of the gun electrodes. The gas valves utilize a pulsed
current in a thin pancake coil to induce image currents in an adjacent
aluminum disk\cite{thomas93} but also employ a plenum for measured,
reproducible gas output. The disk is thus repelled from the coil, creating a
transient opening for gas flow from the plenum into the vacuum chamber.
Restoring force on the disk is provided by a combination of the high pressure
gas in the gas lines and a metal spring. Calibrations indicate that each pulse
injects $10^{20}$--$10^{21}$ hydrogen molecules, depending on the applied
voltage. Optimum timing of gas valve firing is determined empirically by
adjusting the valve firing time to minimize the delay between capacitor bank
trigger and gas breakdown (typically a few $\mu$s). One valve is used for the
eight inner electrode gas injection holes, and two valves are used for the
eight outer electrode holes. Typical gas valve firing voltage is 600~V, with
all three valves injecting on the order of $10^{21}$ hydrogen molecules. It
was found that the amount of gas injected per pulse
slowly increased throughout the day\cite{Yee99}
due to heating in the electrolytic capacitors\cite{Bellan02} powering the gas
valve. A new power supply using metal film polypropylene capacitors has been
constructed recently to eliminate this problem.\cite{Bellan02}

\subsection{Diagnostics}

The images presented in this paper were taken with multiple-frame
charge-coupled device (CCD) cameras. Most of the images were taken with a
Cooke Co.\ HSFC-PRO (Figs.~5, 8, 12, and 14).
The specifications for this camera are as follows: 4 or 8
frames per shot, $1280\times1024$ pixels per frame, 12 bits per pixel,
a 10~ns minimum exposure time, and a minimum inter-frame time of
5~ns (250~ns) in 4 (8) shot mode.  Other images were taken with a DRS Hadland
Imacon 200 (Fig.~11), which has 8 or 16 frames per shot,
$1200\times980$ pixels per
frame, 10 bits per pixel, a minimum 10~ns exposure time,
and a minimum of 5~ns (250~ns) inter-frame time in 8
(16) shot mode. The camera is placed on a tripod and positioned in front of
the right-most vacuum chamber window shown in Fig.~3(top).

Magnetic field data are taken with a radial
array of small pickup coils mounted on a
stainless steel shaft, which can be scanned in $Z$ and rotated in $\phi$.
The array contains twenty groups of three coils
arranged so that all three components of $\mathbf{B}$ are measured from $R=0$
to $R=40$~cm with 2~cm resolution. The coils are commercial chip inductors
(Coil Craft 1008CS-472XGBB) and have calibrated turns $\times$ area $NA
\approx1.3$~cm$^{2}$. The coils are placed into precision-machined slots of a
long thin strip of Delrin, which in turn is slid down a thin-wall stainless
steel tube covered by an alumina tube to prevent metal contact with the plasma
(see Fig.~1). The alumina has an outer diameter of 8.4~mm. The
frequency response, limited by the skin effect of the stainless steel tube, is
good up to 1~MHz. Further details of probe design, construction,
calibration, as well as probe improvements since this work
are reported elsewhere.\cite{Romero03}

A Rogowski coil placed around the ceramic break, which is connected to the
inner electrode, measures total $I_{\mathrm{gun}}$. The Rogowski coil was
hand-wound
with a calibrated $NA = 9.913 \times10^{-3}$~m$^{2}$, and the signal is
integrated via a passive integrator with $RC = 8$~ms. An attenuating Tektronix
P6015 high voltage probe measures $V_{\mathrm{gun}}$.

\subsection{Control and data acquisition}

Triggering of the various power supplies (bias coil, gas valves, and coaxial
gun), the data acquisition system, and the camera is provided by a
programmable 12-channel sequencer with fiber optic outputs. A typical trigger
sequence is as follows: (1)~the bias coil is fired at $t= -10$~ms (due to its
$\approx10$~ms rise-time), (2)~the gas valves are fired at $t= -1.5$~ms (to
accommodate the $\approx1.5$~ms travel time of neutral gas through the gas
lines into the plasma formation region), and (3)~the main capacitor bank
ignitrons are triggered at $t=0$ with gas breakdown typically occuring at $t
\approx4$~$\mu$s.

The digital data acquisition system is a 64-channel VME (Versa Module Europa)
system from Struck Innovative Systems. Each channel has a sampling rate of
105~MHz, a 12-bit resolution, 50~$\Omega$ input impedance, 32~kB memory, and
an input range of -512 to 512~mV with continuously adjustable offset. All
magnetic probe and gun diagnostic signals are digitized on this system. The
digital acquisition system is controlled and the data is displayed and
analyzed using Interactive Data Language (IDL) routines. The magnetic probe
array signals are integrated numerically.

\subsection{Plasma parameters}

The plasmas produced by the coaxial gun have the following nominal 
global parameters:
$n \sim10^{14}$~cm$^{-3}$,\cite{note} $T_{\mathrm{e}} \sim T_{\mathrm{i}}
\sim5$--15~eV,
and $B \approx0.2$--1~kG\@. The global plasma length scale $L$ is
approximately 25~cm, and the typical ion gyroradius is $\rho_{\mathrm{i}}
\approx2$~cm. The characteristic Alfv\'{e}n\ transit time is $\tau_{A}
\approx1.5$~$\mu$s; plasma lifetime is about 15~$\mu$s; resistive diffusion
time is on the order of 1~ms. The Lundquist number $S\equiv\mu_{0} L
V_{\mathrm{A}}/\eta\sim10^{2}$--$10^{3}$, where $V_{\mathrm{A}}$ is the
Alfv\'{e}n\ speed and $\eta$ is the classical resistivity. Thus, the 15~$\mu$s
duration plasma dynamics lasts several Alfv\'{e}n\ times, and the magnetic
flux is reasonably frozen into the plasma.

\section{Experimental results}

\label{results}

\subsection{Plasma breakdown and initial evolution}

The Paschen condition for gas breakdown\cite{vonengel65,bellan00} shows that
the voltage required for gas breakdown between two electrodes separated by
distance $d$ depends on the product $pd$, where $p$ is the pressure in the
region between electrodes. The dependence is such that the breakdown voltage
becomes infinite when the $pd$ product becomes less than about half the value
at which the breakdown voltage is minimum. This property is exploited using a
fast, localized gas injection, as shown in Fig.~3(bottom).
The gas valve operation is
adjusted so that $pd$ satisfies the Paschen condition along the bias field
lines. Because of the 6~mm size of the inter-electrode gap and
the negligible gas pressure in the gap, the $pd$ value in the gap is so small
as to be well to the left of the Paschen minimum such that breakdown cannot
occur there. This setup demonstrates the counter-intuitive fact that
a very small distance between conductors in vacuum can provide perfect
electrical insulation.\cite{braun88}
This fact was also utilized in the Caltech solar prominence experimental
setup.\cite{Bellan98}

Breakdown along the bias field lines is shown in frame~1 of Fig.~5,
which is a CCD camera image of the plasma within approximately the first
microsecond after breakdown. The electrodes are located in the right hand side
of each frame; the circular gap between electrodes is identified by an arrow.
The locations of the eight bright legs
correspond to the locations of the eight pairs of gas injection holes, and the
shape of the legs matches the calculated bias field configuration. These
observations indicate that breakdown is occurring along the bias field lines
and that the gas pressure is non-uniform and concentrated azimuthally at the
eight $\phi$ positions corresponding to the gas injection holes. The breakdown
occurs approximately 4--5~$\mu$s after high-voltage is applied (see
Fig.~4). In order to minimize jitter and maximize reproducibility,
the gas injection timing was adjusted to minimize the time delay before
breakdown.

The gun current ramps up steeply during the next 10~$\mu$s (see
Fig.~4), and the eight bright legs expand and merge, as shown in
the next two frames of Fig.~5. The rising gun current corresponds to
an increase in toroidal magnetic flux linking the bias field lines,
and therefore magnetic helicity is injected at the rate $2V_{\mathrm{gun}}%
\psi_{\mathrm{gun}}$.\cite{bellan00} The applied gun voltage creates a radial
electric field. From the radial component of the ideal Ohm's law,
\begin{equation}
E_{\rm R} + U_\phi B_{\rm Z} - U_{\rm Z} B_\phi = 0,
\end{equation}
it is seen that a combination of toroidal rotation $U_{\phi}$ and
axial flow $U_{\rm Z}$ will arise to balance $E_{\mathrm{R}}$.
Note that an astrophysical accretion disk is similar to a coaxial gun
but with the driving term coming from the Keplerian disk rotation $U_{\phi}$
which gives rise to $E_{\mathrm{R}}$.\cite{hsu02b} As described below, further
evolution of the plasma after frame~3 of Fig.~5 is dependent on the
peak value of $\lambda_{\mathrm{gun}}=\mu_{0}I_{\mathrm{gun}}/\psi
_{\mathrm{gun}}$.

\subsection{Plasma morphologies resulting from $\lambda_{\mathrm{gun}}$
parameter scan}

After the initial formation stage, the plasma can evolve into three distinct
morphologies depending on peak $\lambda_{\mathrm{gun}}$.\cite{hsu02b} Similar
regimes were also observed in previous
Caltech experiments using a conventional cylindrical coaxial gun.\cite{yee00}
The parameter $\lambda_{\mathrm{gun}}$ can be thought of as a boundary
condition imposed on the plasma at the gun surface. Its functional form
$\lambda_{\mathrm{gun}} = \mu_{0} I_{\mathrm{gun}}/\psi_{\mathrm{gun}}$ can be
easily derived by integrating $\nabla\times\mathbf{B} = \lambda\mathbf{B}$
over the inner gun electrode surface. The parameter $\lambda_{\mathrm{gun}}$
is varied experimentally by adjusting the main capacitor bank voltage (which
controls $I_{\mathrm{gun}}$) and the bias coil bank voltage (which controls
$\psi_{\mathrm{gun}}$). A plot of $\lambda_{\mathrm{gun}}$ parameter space is
shown in Fig.~6, with the different observed plasma morphologies (I, II, III)
indicated.

\subsubsection{Formation of stable plasma column at low $\lambda
_{\mathrm{gun}}$}

The formation of a central plasma column along the $Z$ direction
is observed for values of
$\lambda_{\mathrm{gun}} \lesssim40$~m$^{-1}$, as shown in Fig.~5.
Magnetic probe measurements in this regime (Fig.~7) show that the
column has magnetic field radial profiles resembling a screw pinch. The axial
expansion of the column is derived from Fig.~5 to be approximately
40~km/s which is of the order of the Alfv\'{e}n\ speed. The filamentary shape
of this column provides compelling experimental evidence for
magnetically-driven models of astrophysical jet
collimation,\cite{lynden-bell96,meier01,ouyed03,bellan03}
in which an ionized rotating accretion disk winds up
a background magnetic field and injects magnetic helicity into the
disk corona.

\subsubsection{Kink instability of central column at intermediate
$\lambda_{\mathrm{gun}}$}

For $40$~m$^{-1} \lesssim\lambda_{\mathrm{gun}} \lesssim60$~m$^{-1}$, the
plasma column forms along the $Z$ direction but then develops a helical
instability with toroidal mode number $n=1$, as shown in Fig.~8. The
helical instability is shown to be consistent with an ideal MHD kink
instability by two independent experimental measurements described below.

The Kruskal-Shafranov condition for MHD kink instability in cylindrical
geometry is
\begin{equation}
q(a) = 2\pi a B_{\mathrm{Z}}(a)/LB_{\phi}(a) < 1,
\label{ks-condition}
\end{equation}
where $q$ is the safety factor and
$a$ and $L$ are the column radius and length, respectively. By invoking
the relationships
$\psi_{\mathrm{gun}} \approx\pi a^{2} B_{\mathrm{Z}}(a)$ and $B_{\phi}(a) =
\mu_{0} I_{\mathrm{gun}}/ 2\pi a = \lambda_{\mathrm{gun}} \psi_{\mathrm{gun}}
/ a$, the Kruskal-Shafranov condition for kink instability can be expressed
as\cite{duck97, hsu02b}
\begin{equation}
\lambda_{\mathrm{gun}} > 4\pi/L.
\end{equation}
Thus, the experimentally observed instability onset can be compared to the
Kruskal-Shafranov condition by knowing only $\lambda_{\mathrm{gun}}$ and $L$.
Figure~9 shows a scatter plot of $\lambda_{\mathrm{gun}}$
versus $L$ for an ensemble of shots in which a plasma column forms. The data
points represent both different shots and different times within the same shot
(\emph{i.e.}, before and after the instability occurs). It is seen that the
threshold for the appearance of the helical instability is in good agreement
with the Kruskal-Shafranov limit (dashed line in Fig.~9).

Consistency with the Kruskal-Shafranov limit was also checked with direct
magnetic probe measurements. Radial profiles of $B_{\mathrm{Z}}$ and $B_{\phi
}$ yielded $q$ profiles as a function of time, as shown in
Fig.~10. It is shown that $q$ drops to unity around 10.5~$\mu$s.
The corresponding image sequence for this shot, shown in Fig.~11,
indicates that the helical instability develops at this same time. Thus the
observed helical instability is consistent with an ideal kink, with onset
occurring when $q$ drops to unity.  The $q$ is initially greater
than unity because the bias poloidal flux dominates and $L$ is
small.  Then $L$ increases quickly and $\Bt$ also increases to a 
lesser degree, and therefore
$q$ drops according to Eq.~(\ref{ks-condition}).

The growth of the kink mode was captured by shortening the inter-frame time of
the CCD camera to 0.25~$\mu$s (see Fig.~12). The mode amplitude as a
function of time is shown in Fig.~13 and is seen to exhibit a
fairly linear growth rate.

\subsubsection{Quick plasma detachment at high $\lambda_{\mathrm{gun}}$}

At values of $\lambda_{\mathrm{gun}} \gtrsim60$~m$^{-1}$, the plasma appears
to ``pinch-off'' without the formation of a central column, as shown in
Fig.~14. From the image sequence, the propagation velocity of the
detached plasma along the $Z$-axis is calculated to be around 50~km/s (of the
order of the Alfv\'{e}n\ speed). Magnetic probe measurements indicate that the
detached plasma has much stronger toroidal than poloidal field
(Fig.~15).  It is possible
that this toroidal-flux-rich detached plasma will relax to a spheromak
configuration with equivalent toroidal and poloidal flux. 

Poloidal flux amplification, as measured by the magnetic
probe array, is large in this regime.  Flux amplification
for all three regimes is shown
in Fig.~16 as a function of peak $\lambda_{\mathrm{gun}}$.
In the kinked regime, the flux
amplification is typically around a factor of two, and the flux amplification
mechanism in this regime (discussed below) has been identified. The flux
amplification is neligible in the low $\lambda_{\mathrm{gun}}$ regime, but it
is very large in the high $\lambda_{\mathrm{gun}}$ regime. The mechanism for
the high $\lambda_{\mathrm{gun}}$ regime is not known, but it is probably
related to the mechanism operating in the kink regime.  An interesting
question requiring further study is what determintes
the maximum amount of flux amplification.
The next section shows
that the kink is a mechanism for poloidal flux amplification and spheromak
formation. 

\subsection{Kink as mechanism for poloidal flux amplification and spheromak
formation}

\label{spheromak-formation}

The kink instability is identified experimentally as a mechanism for
(1)~poloidal flux amplification and (2)~spheromak formation.\cite{hsu03}
Both are experimentally
observed immediately following kinking of the central column. This can be seen
by examining the evolution of poloidal flux contours and maximum $\psi$
(Fig.~17)
calculated from time-resolved magnetic probe data and the
sequence of CCD images showing the kink (Fig.~11).

\subsubsection{Paramagnetism of the kink}

The kink modifies the direction of current flow from purely
$Z$ (poloidal) to a partially $\phi$ (toroidal) direction.
Consequently, it converts toroidal to poloidal flux. This process
is paramagnetic, amplifying $\psi$ over the initial applied $\psi
_{\mathrm{gun}} = 1.7$~mWb. The paramagnetism is understood by realizing that
kinks involve perturbations with dependence $\mathrm{exp}(i\mathbf{k \cdot
x})$ where $\mathbf{k \cdot B}=0$.
The latter means
\begin{equation}
k_{\mathrm{Z}} = -k_\phi B_{\phi}/B_{\mathrm{Z}},
\label{kz}
\end{equation}
where $k_{\rm R}=0$.
The trajectory of the kink
is determined by considering a locus of constant phase of the
perturbation,
{\em i.e.}, $\mathbf{k}\cdot \mathbf{x}={\rm constant}$, such that
\begin{equation}
k_Z Z + k_\phi R\phi = {\rm constant}.
\label{phase}
\end{equation}
Thus, substituting Eq.~(\ref{kz}) into Eq.~(\ref{phase})
gives the coordinates of the helix,
$\phi = (B_{\phi}/R B_{\mathrm{Z}})Z + {\rm constant}$,
from which it can be seen that the kinked current
channel is a right-handed helix if $J_{\mathrm{Z}} B_{\mathrm{Z}}
\sim \Bt \Bz > 0$ and a left-handed helix if $J_{\rm Z} B_{\rm Z} < 0$.
Thus, the helix will always have the form of a solenoid with
the appropriate handedness to amplify the original $B_{\mathrm{Z}}$. The
additional $\psi$ introduced by the kink can be estimated by considering the
helically deformed current channel to be a solenoid with current $I$, turns
per length $1/L$, and radius $a$. The solenoid formula for the field inside
the solenoid is $B_{\mathrm{Z}}=\mu_{0} I/L$; the $\psi$ produced by the
solenoid is $\pi a^{2} B_{\mathrm{Z}}$ and thus depends non-linearly on the
kink amplitude $a$. Using the measured values $a\approx5$~cm, $L \approx
20$~cm, and $I\approx60$~kA at 13.5~$\mu$s, the $\psi$ generated by the kink
is predicted to be approximately 1~mWb, which is within a factor of 2 of the
observed amplification of $\psi_{\mathrm{max}}$ over $\psi_{\mathrm{gun}}$.
The discrepancy is within the accuracy of $a$ and $L$ measurements and of the
$\psi$ calculation assuming axisymmetry in the presence of the rotating kink.
Because the coaxial gun injects only toroidal flux, 3D plasma dynamics must
be responsible for $\psi$ exceeding $\psi_{\mathrm{gun}}$. The dynamics are
provided by the kink, which is the mechanism by which toroidal flux is
converted to poloidal flux.

\subsubsection{Evidence for spheromak formation}

The appearance of the kink is followed immediately by three signatures of
spheromak formation: (1)~appearance of closed $\psi$-contours [calculated
assuming axisymmetry, $\psi(R,t) = \int_{0}^{R} 2 \pi R^{\prime}B_{\mathrm{Z}%
}(R^{\prime},t)\, \mathrm{d} R^{\prime}$], (2)~$\psi$-amplification, and
(3)~magnetic field radial profiles consistent with spheromak formation. It is
important to note that the relationship between closed $\psi$-contours and
closed flux surfaces becomes ambiguous when axisymmetry is broken. Thus, in
the presence of a non-axisymmetric kink, closed $\psi$-contours indicate
closed flux surfaces only in a time-averaged manner. Shortly after 13~$\mu$s,
the kink breaks apart (Fig.~11). This coincides with signatures of
spheromak formation as observed in the magnetic probe measurements
(Fig.~17). At approximately 13~$\mu$s, closed $\psi$-contours appear
and $\psi_{\mathrm{max}}$ is amplified to larger than $\psi_{\mathrm{gun}}$
($\psi$-amplification is due mainly to broadening of the $B_{\mathrm{Z}}$
profile after 12~$\mu$s). At 15~$\mu$s, magnetic field profiles consistent
with spheromak formation are observed (Fig.~18). The measured
radial profiles of $B_{\mathrm{Z}}$ and $B_{\phi}$ at 15~$\mu$s are compared
with Taylor state solutions\cite{taylor86} in cylindrical geometry,
\emph{i.e.}\ uniform $\lambda$ solutions of $\nabla\times\mathbf{B} =
\lambda\mathbf{B}.$ The solutions are $B_{\mathrm{Z}} \sim J_{0}(\lambda R)$
and $B_{\phi}\sim J_{1}(\lambda R)$, where $J_{0}$ and $J_{1}$ are Bessel
functions of order zero and one, respectively, and the best fit is found for
$\lambda\approx15$~m$^{-1}$ with a radial offset of 4~cm (displacement of
spheromak off the geometric axis). The slight disagreement between measured
profiles and the Taylor solution is not surprising since the spheromak is
expected to be either (1)~still undergoing relaxation toward the Taylor state
or (2)~in a modified relaxed state since it is still being driven by the gun,
where peak $\lambda_{\mathrm{gun}} \approx50$~m$^{-1}$.

\section{Discussion}

\label{discussion}

\subsection{Spheromak formation and topology}

Spheromak formation involves a sequence of steps that in principle should be
possible to visualize mentally. However, the inherent geometric and
topological complexity of these steps has made such a visualization so
challenging that no consensus exists on the precise form of this sequence.
This shortcoming is important because without being able to visualize how the
3D topology evolves during the spheromak formation sequence, it is difficult
to optimize this process and the closely related sustainment
process.\cite{woodruff03} Because visualization of the process has proved
difficult, the main appoach used to date for experimentally
characterizing spheromak
formation and sustainment is modal analysis, a method where Fourier modes are
first identified and then the time dependence and spatial profile of these
modes are measured and compared with theoretical models. While useful in many
respects, modal analysis gives little insight into 3D topological issues such
as linkage, connectivity, knottedness, and conversion between toroidal flux
and poloidal flux. The photographs of evolving magnetized plasmas in the
present experiments have provided an actual visualization which helps provide
new insights into these issues. New interpretations are given here and placed
into context with respect to past experimental
work.\cite{lindberg61, knox86, browning92,
nagata93,duck97,willet99}  It should also be noted that
many aspects of the present work are qualitatively similar to
recent time-dependent MHD simulations of spheromak formation
via electrostatic helicity injection.\cite{Sovinec01}

Bounded, axisymmetric, force-free plasma equilibria were first discussed in
the 1950's.\cite{woltjer58a,woltjer58b, chandrasekhar56,chandrasekhar57} These
theoretical models invoked rather abstract astrophysical contexts and did not
contain descriptions of the detailed processes by which such equilibria could
be created. A hypothetical analogy to this situation would be to be in
possesion of an abstract theoretical model demonstrating that soap bubbles
should in principle exist, while never having observed a soap bubble nor being
aware of how to blow soap bubbles. The soap bubble analogy is apt because a
bounded, axisymmetric, force-free equilibrium can be thought of as being a
bubble (local maximum) of poloidal flux $\psi(R,Z)$. First derivatives
(gradients) of $\psi$ show that a local $\psi(R,Z)$ maximum constitutes a
magnetic axis, and second derivatives (Laplacian-like operators) of $\psi$
show that a toroidal current necessarily flows along this magnetic axis.

Lacking detailed knowledge for how to make these ``magnetic bubbles'' is a
non-trivial shortcoming because Cowling's anti-dynamo theorem\cite{cowling34}
shows that there is no simple way to make such $\psi$ bubbles. Specifically,
Cowling showed that creation/sustainment of an axisymmetric, steady-state
toroidal current cannot be achieved via axisymmetric (\emph{i.e.}, simple)
means. Since creation of an axisymmetric toroidal current is tantamount to
creating a local $\psi(R,Z)$ maximum, the Cowling anti-dynamo theorem
effectively requires that any poloidal flux amplification process must be
non-axisymmetric. Thus, spheromak formation and sustainment necessarily involve
non-axisymmetric processes.

Helical (\emph{i.e.}, non-axisymmetric) distortions of magnetized coaxial gun
central columns and associated poloidal flux amplification were first observed
in the early 1960's in experiments by Alfven et al.\cite{alfven60} and
Lindberg et al.\cite{lindberg61} It is interesting to note that these early
magnetized coaxial gun experiments revealed phenomena relating to spheromak
formation nearly two decades before the word ``spheromak'' was coined by
Rosenbluth and Bussac.\cite{rosenbluth79} In particular, Lindberg et
al.\cite{lindberg61} imaged the helical instability of an initially
azimuthally symmetric axial current channel, characterized this as an ``$m=1$
instability,'' noted that ``such instabilities are well known in the theories
of pinches,'' and measured an amplification of the poloidal flux. Lindberg et
al.\ interpreted this central column helix as a kink instability of the
\emph{geometric axis} current and postulated that the helical deformation of
the current path was responsible for observed poloidal flux amplification.
However, Lindberg et al.\ did not provide quantitative measurements
establishing that the observed helical instability was indeed a kink
(\emph{i.e.}, that the configuration destabilized when the $q=1$ kink
stability threshold was crossed) nor that the instability produced a magnitude
and polarity of poloidal flux consistent with observation.

Similar modes (now called $n=1$ modes) were also observed in all later coaxial
gun experiments,\cite{knox86, browning92, nagata93, jarboe94}
but there have been two
distinct, differing interpretations of these modes. In the first
interpretation, referred to here as ``rotating static
mode,'' the modes are assumed to have a steady-state geometry
in some rest frame so that their apparent laboratory frame time-dependence is
solely due to the rotation of the mode rest frame with respect to the
laboratory
frame. In this interpretation the modes are thus assumed to have a laboratory
frame time-dependence $\sim\cos(n\theta-\omega t)$ so that an observer located
in a frame rotating with the mode (\emph{i.e.}, a frame with coordinate
$\theta^{\prime}=\theta-\omega t/n$) would see a time-independent, or static,
mode. In the second interpretation, referred to here as 
``relaxation oscillation,'' the modes are not static in any
frame but, rather, are growing modes which start with zero amplitude, build up
to finite amplitude, undergo a nonlinear crash, and then the process repeats.
There might be rotation as well, but even if one moved to an optimally chosen
rotating frame, there still would be relaxation oscillations. Thus, in the
relaxation oscillation case, the mode time-dependence during its growth stage
would be $\sim e^{\gamma t}\cos n\theta$ or possibly $\sim e^{\gamma t}%
\cos(n\theta-\omega t)$ so that no change of observer rest frame would cause
the mode to appear static.

The $n=1$ modes were typically observed as oscillations in magnetic field
during the \emph{sustainment} (as opposed to formation) phase. However, it
seems likely that sustainment ought to involve much the same processes as
formation since formation requires creating a localized poloidal flux maximum,
and sustainment requires maintaining this maximum against resistive losses.
Knox et al.\cite{knox86} observed $n=1$ and higher order modes using wall
magnetic probe arrays and proposed that sustainment was due to a rotating
static mode which they called a ``rotating internal kink distortion.'' On the
other hand Nagata et al.\cite{nagata93} made similar magnetic probe
measurements but interpreted sustainment as a sequence of relaxation
oscillations (``collapse and recovery"). Thus Knox et al.\ considered the mode
as a rotating perturbation of an otherwise axisymmetric flux surface, whereas
Nagata et al.\ interpreted the mode as being the creation and destruction of
flux surfaces.

The SPHEX group\cite{browning92, duck97, willet99} used internal
probes to identify two distinct structures in their experiment, namely (i)~a
central column of open magnetic field lines with a force-free current flowing
from the gun electrode along the geometric axis and (ii)~an annulus (torus) of
closed field lines and toroidal current linking the central column. The SPHEX
group found that the $n=1$ oscillations were primarily in the central column
and, using probes, showed that the central column had a helical deformation.
The helical deformation was inferred on the assumption that the $n=1$
oscillation was a rotating static mode, and this mode was interpreted to be a
nonlinearly saturated kink instability of the \emph{geometric axis} current
rotating toroidally at the $E\times B$ frequency.\cite{brennan99}
It is important to emphasize
that the rotation was \emph{assumed}\cite{duck97} but not directly measured.
Because the signal was observed to be periodic, it was assumed that the
fluctuations were of the form $\cos(\theta-\omega t)$ so that observations
made at different times could be used to map out the spatial profile of the
assumed rotating structure. However, this assumption of rotation 
misses the interpretation that there is a relaxation oscillation with
repeated births of new helices, \emph{i.e.}, the rotation assumption 
did not consider
the possibility of a repetitive sequence of modes with time behavior
$e^{\gamma t}\cos\theta$.

The present results suggest that the $n=1$ fluctuations result from repeated
births of new kinked helices with time dependence $\sim e^{\gamma t}\cos
\theta$ and that these helices detach and merge into the time-averaged closed
flux region of a spheromak. According to this interpretation of spheromak
formation/sustainment, the current starts out aligned along the geometric axis
and so is initially purely poloidal. Then, upon kinking (\emph{i.e.}, the
current channel becoming helical so that it is effectively a solenoid), the
current channel axis has a toroidal component which creates new poloidal flux.
Kink growth means that the radius of the helix increases. Growth of the kink,
in turn, means that the radius of the helix increases until this radius
becomes so large that the helical current channel merges with any existing
spheromak toroidal current, thereby amplifying/sustaining the poloidal flux.
New current will now flow from the gun along a new straight central column
aligned along the geometric axis, and initially, this new central current
column will be stable against kinking because of the strong axial magnetic
field resulting from the strong toroidal current flowing in the spheromak.
However, because no electromotive force sustains the spheromak toroidal
current against Ohmic decay, the spheromak toroidal current decays with time
so that the poloidal field on the geometric axis becomes weaker. The open
poloidal field on the geometric axis weakens to the point that a new kink
develops on the geometric axis. The new kink deforms into a new solenoid which
merges with the spheromak, replenishing the toroidal curent in the spheromak,
and increasing the geometric axis open poloidal field so that the process
starts over again. Thus, power flow from the geometric axis region to the
spheromak is not the result of rotation of a saturated kink but is instead due
to the repetitive creation of successive helical (kinked) geometric
axis current channels
which expand nonlinearly in radius and merge with the spheromak.

The interpretation in the above paragraph is buttressed by considering what
happens when a simple solenoid, consisting of many turns of wire, is rotated
about its own axis. Such a rotation does not result in flow of magnetic
energy or in transformation of toroidal field energy into poloidal field
energy. In fact rotation of a solenoid about its axis does not change the
magnetic field produced by the solenoid at all because rotation of a solenoid
does not change the relative toroidal drift velocity between electrons and
ions in the solenoid wire. In contrast, increasing the radius of the solenoid,
as would be the consequence of a growing kink instability, increases the
poloidal flux through the bore of the solenoid. If the conductor forming the
solenoid (\emph{i.e.}, the initial geometric axis current channel) is a flux
conserver, then this increase of poloidal flux linking the conductor must be
at the expense of a decrease in the toroidal flux linking the conductor,
resulting in a conversion from toroidal to poloidal flux due to the kink.

The primary result in this work regarding spheromak formation is the
experimental demonstration that the kink instability of the geometric axis
current and its growth result in the poloidal flux amplification required to
produce the local maximum of poloidal flux necessary for spheromak formation.
This result is qualitatively similar to the point of view expressed by
Lindberg et al.\ and Nagata et al.\ but differs significantly from the point
of view expressed by Knox et al.\ and the SPHEX group. The present work has
provided a quantitative demonstration that the onset of the helical
instability is consistent with the Kruskal-Shafranov condition for kink
instability and a theoretical demonstration that the kink is necessarily
paramagnetic (\emph{i.e.}, amplifies poloidal flux). Furthermore, it was shown
that the poloidal flux amplification calculated using the visually observed
kink dimensions are consistent with the magnetic probe measurements of
poloidal flux. These measurements support the interpretation that
poloidal
flux amplification is due to kink \emph{growth} and not kink rotation.
Repetition of this sequence with successive, new kinked current channels
during sustainment could explain the previously observed $n=1$ fluctuations.
The flux amplification mechanism would be insensitive to the magnitude or
direction of the rotation frequency.

\subsection{Astrophysical jets}

Magnetically-driven astrophysical jets are intriguing astronomical
phenomena which have been
observed for nearly a century. These jets are highly collimated and have been
observed to emanate from stars, black holes, and active galactic nuclei; the
sources are generically called central objects. The characteristic size of
jets ranges over many orders of magnitude, but in all cases the axial extent
of the jet is much larger than the characteristic dimension of the source. 
Theoretical models of jet formation and collimation have
been discussed at length in the literature.\cite{begelman84,blandford00,
meier01}  Observations of helical magnetic
fields in jets\cite{asada02} provide especially tantalizing evidence that
plasma physics plays an important role governing
jet collimation, structure, and dynamics.  Theoretical models
based on MHD have become the leading candidates to explain jet
formation and collimation.  Many of these models show how
axisymmetric force-free or nearly force-free plasma equilibria could be
made to resemble the long thin morphology of a jet.\cite{lynden-bell96, li01,
lovelace02}  More recently, time-dependent
3D MHD simulations\cite{meier01,ouyed03} have illustrated dynamical
collimation
starting with an initial ``seed'' magnetic field configuration.  Despite
impressive similarities between the theoretical models and
astronomical observations of jets, experiments to
test the underpinnings of these theories have been lacking.

The planar spheromak gun formation experiment discussed here has a geometry
very similar to the geometry associated with the formation region of
magnetically-driven astrophysical jets (Fig.~2), and so this work can be
considered as a laboratory investigation into the underlying processes
governing astrophysical jet formation, collimation, and
acceleration.\cite{hsu02b} 
The MHD theories are based on the generation of an axisymmetric poloidal
current with geometry essentially the same as the poloidal current in the
central column of the present experiments. In the astrophysical situation,
this poloidal current is driven by a radial electric field resulting from the
rotation of an accretion disk located in the equatorial plane of the central
object. The rotation of the accretion disk cuts across a pre-existing poloidal
magnetic field, resulting in an electric field $\mathbf{E=-}U_{\phi}\hat{\phi
}\times B_{Z}\hat{Z}=-U_{\phi}B_{Z}\hat{R}$, which is equivalent to the radial
electric field applied across the gap between inner and outer electrodes in
the experiment.  
In both cases, magnetic helicity is injected
at the rate ${\rm d}K/{\rm d}t = 2 V \psi$, where $V$ is the potential
drop between the inner and outer electrodes (or between star and disk)
and $\psi$ is the poloidal magnetic flux linking the two.

The present work demonstrates experimentally for the first time that
magnetic helicity injection in relevant geometry leads to the formation
and collimation of a jet (Fig.~5).  In particular, the data
indicate both a jet-like flow and a collimation of this flow,
\emph{i.e.}, the lengthening central column is the jet, and the narrowing of
the central column is the collimation. The jet flow is consistent with a
recent model\cite{bellan03} proposing that axial gradients in $B_{\phi}^{2}$
drive axial flows from where the current channel radius is small to where the
channel radius is large and that collimation occurs where $\nabla
\cdot\mathbf{U}$ is negative so that there is a compression of the toroidal
flux frozen into the flow.  The present work also
shows that collimation can occur within just one
twist of the poloidal magnetic field, as demonstrated by the fact
that the kink instability occurs when $q$ drops to unity, {\em
after} collimation has already been observed.
It is interesting to note that the kink instability has been
proposed to explain the observed knotted
structures in astrophysical jets.\cite{nakamura01}

\section{Summary}

\label{summary}

Magnetic helicity injection experiments using a simple magnetized coaxial gun
setup have revealed insights into the dynamics of spheromak formation and
collimation of magnetically-driven astrophysical jets. Depending on the peak
value of $\lambda_{\mathrm{gun}}$, three different plasma morphologies were
observed: (1)~a collimated jet at $\lambda_{\mathrm{gun}} \lesssim40$~m$^{-1}%
$, (2)~kink instability of the jet at $40 \lesssim\lambda_{\mathrm{gun}}
\lesssim60$~m$^{-1}$, and (3)~a quickly detached plasma at $\lambda
_{\mathrm{gun}} \gtrsim 60$~m$^{-1}$. The kink instability was shown to provide
the mechanism for poloidal flux amplification in coaxial gun spheromak
formation.
The latter result supports a model in which repeated births of kinked helices
can form and sustain spheromaks. It is proposed that this mechanism, rather
than a rotating static saturated kink which does
not lead to flux amplification, could explain past observations of
$n=1$ fluctuations in coaxial gun spheromak experiments.  
A collimated jet is formed by driving electric current along a
background poloidal field in a geometry relevant for
magnetically-driven astrophysical jets.  
Collimation is observed well before kink onset.  Since kink
onset takes place when the magnetic field has undergone one
complete twist, consistent with the Kruskal-Shafranov theory,
these observations demonstrate that collimation occurs before
the initial poloidal field has undergone one complete twist.
Kink dynamics could be related to wiggled structures in astronomical
observations of galactic jets.

\acknowledgments{This work was supported by a
U.S.\ Dept.\ of Energy (DoE) Fusion Energy Sciences Postdoctoral Fellowship
and DoE grant no.~DE-FG03-98ER544561.
The authors acknowledge S.~Pracko, C.~Romero-Talam\'{a}s,
F.~Cosso, and D.~Felt for technical assistance and also Dr.~S.~Woodruff for
discussions about past work on spheromak
$n=1$ central column instabilities.}



\newpage

\noindent Figure captions\\

\noindent Figure 1.  End-view photograph of
coaxial gun and rotatable magnetic probe array (end-view).\\

\noindent Figure 2.  Schematic of essential ingredients of both a coaxial
gun and astrophysical star-disk system.\\

\noindent Figure 3.  Top: large vacuum
chamber with multiple diagnostic ports and coaxial gun mounted on right
end-dome. Bottom: Side-view schematic of planar coaxial gun, showing inner
electrode (blue), outer electrode (green), gas feed lines, contours of
constant bias poloidal flux, and cylindrical coordinate system.\\

\noindent Figure 4. Typical $I_{\mathrm{gun}}$ (solid) $V_{\mathrm{gun}}$
(dashed) traces. Breakdown occurs at
$t=2$-3~$\mu$s when $V_{\mathrm{gun}}$ drops and $I_{\mathrm{gun}}$ rises.\\

\noindent Figure 5.  CCD images
of plasma evolution (shot 1210) for low $\lambda_{\mathrm{gun}}$ regime, in
which a central plasma column forms.  The circular gap between electrodes
is indicated by an arrow.\\

\noindent Figure 6.  Plot of $\lambda
_{\mathrm{gun}}=\mu_{0}I_{\mathrm{gun}}/\psi_{\mathrm{gun}}$ parameter space
showing the three plasma regimes.\\

\noindent Figure 7.  Magnetic field and total
current as a function of $R$ for low $\lgun$ regime central column
(shot 2465).\\

\noindent Figure 8.  CCD images
of plasma evolution (shot 1233) for intermediate $\lambda_{\mathrm{gun}}$
kinked plasma regime.\\

\noindent Figure 9.  Plot of $\lambda
_{\mathrm{gun}}$ versus column length $L$, showing consistency with
Kruskal-Shafranov condition $\lambda_{\mathrm{gun}} > 4\pi/L$ for MHD kink
instability.\\

\noindent Figure 10.  Radial profiles of
$q=2\pi R B_{\mathrm{Z}}/ L B_{\phi}$ (shot 2472). The column develops a
helical instability at the same time as $q$ near the axis drops to unity, also
showing consistency with Kruskal-Shafranov condition for MHD kink
instability.\\

\noindent Figure 11.  Images of kink development (shot 2472).\\

\noindent Figure 12.  High
time-resolution images of kink growth (shot 1247). Inter-frame time is
250~ns.\\

\noindent Figure 13.  Mode amplitude of
kink perturbation versus time.\\

\noindent Figure 14.  Images of
plasma evolution at high $\lambda_{\mathrm{gun}}$ (shot 1181).\\

\noindent Figure 15.  Radial profiles of $B_{\rm tor}$ and $B_{\rm pol}$ for
high $\lgun$ plasma at $t=12$~$\mu$s (shot 2457).\\

\noindent Figure 16.  Flux
amplification versus peak $\lambda_{\mathrm{gun}}$.\\

\noindent Figure 17.  Top: poloidal magnetic
field vectors and $\psi$ contours as a function of $R$ and $t$. Bottom: Time
evolution of $\psi$ at $R=19$~cm (shot 2472). Flux amplification is observed
during time of the kink and appearance of closed $\psi$ contours.\\

\noindent Figure 18.  Comparison of
$B_{\mathrm{Z}}$ (diamonds) and $B_{\phi}$ (squares) profiles at $t=15$~$\mu$s
with Taylor state (solid line). Profile location is indicated by vertical
dashed line in top panel of Fig.~17.

\newpage

\begin{figure}[ptb]
\includegraphics[width=3truein]{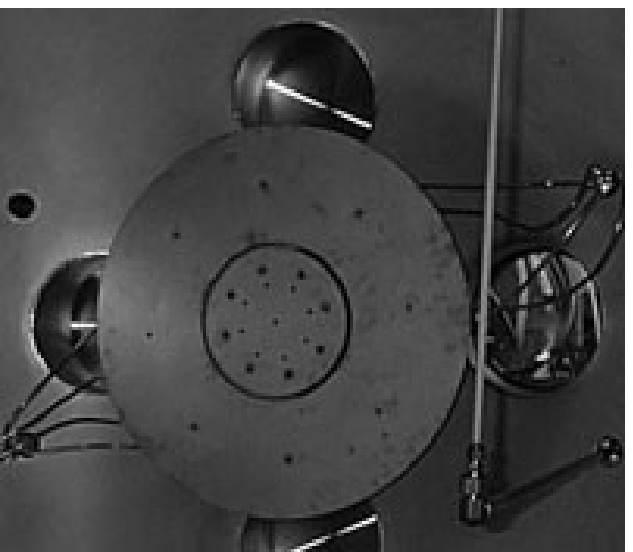}\\
Figure 1
\label{gun-photo}%
\end{figure}

\begin{figure}[ptb]
\includegraphics[width=3.5truein]{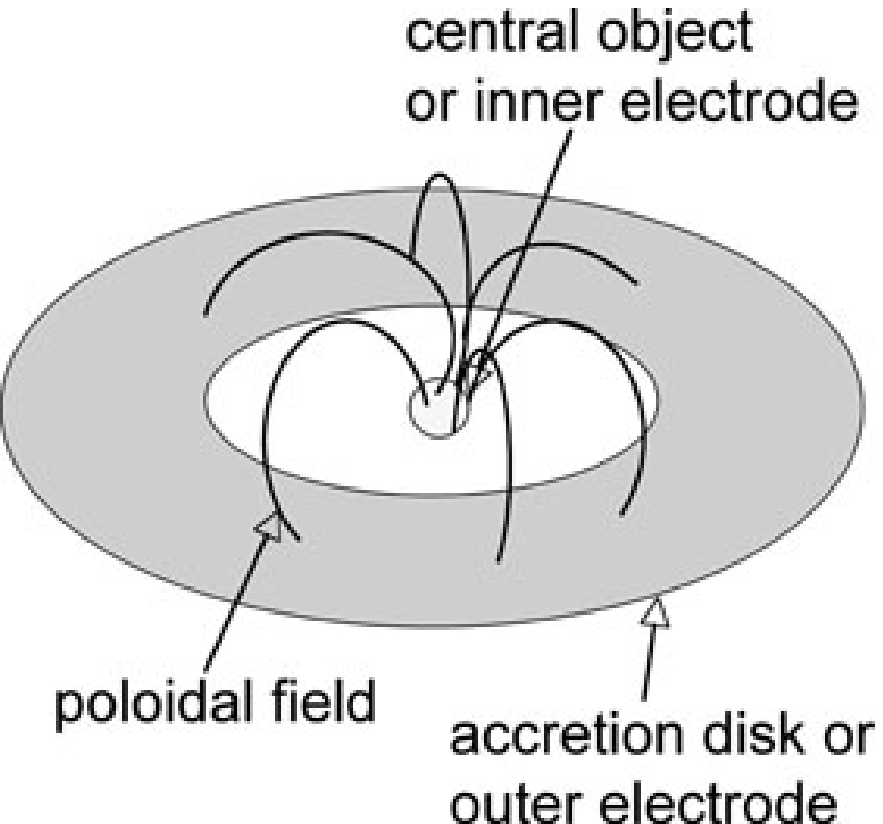}\\
Figure 2
\label{jet-exp}%
\end{figure}

\begin{figure}[ptb]
\includegraphics[width=3truein]{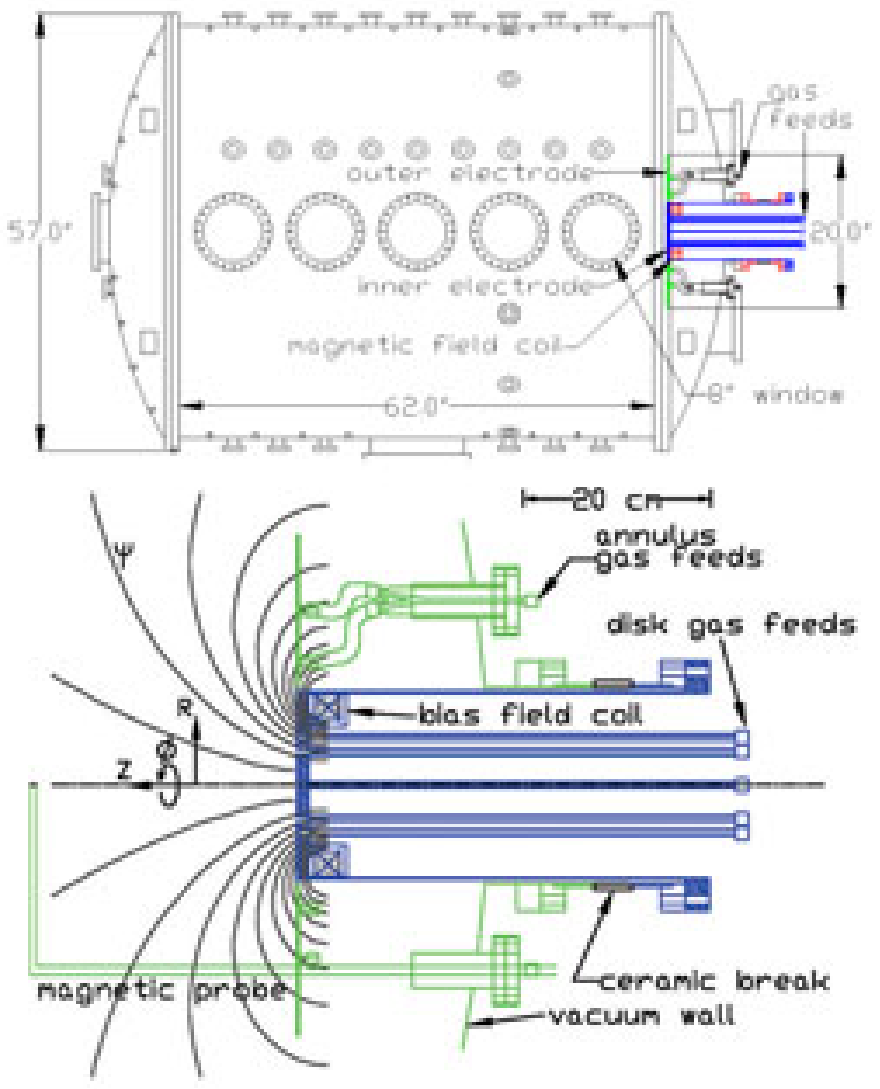}\\
Figure 3
\label{chamber-gun}%
\end{figure}

\begin{figure}[ptb]
\includegraphics[width=3truein]{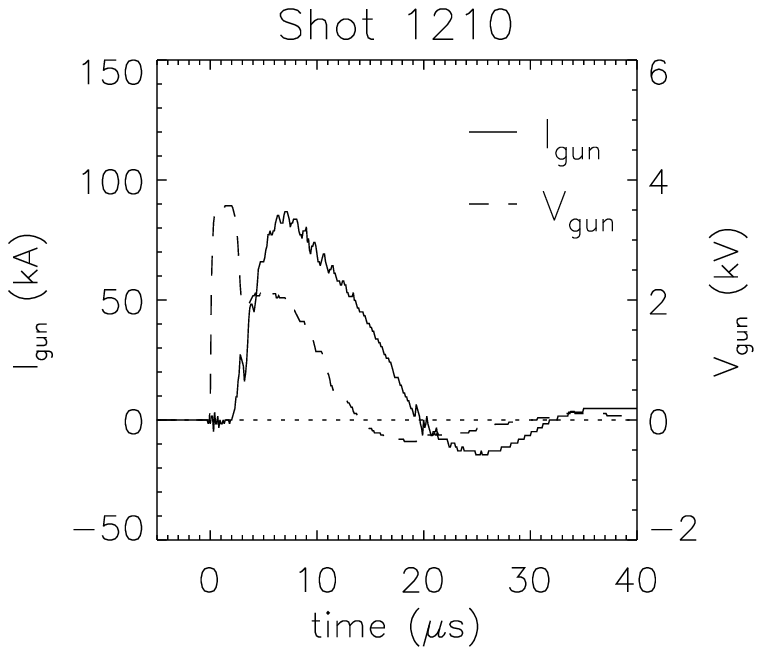}\\
Figure 4
\label{1210iv}%
\end{figure}

\begin{figure}[ptb]
\includegraphics[width=3truein]{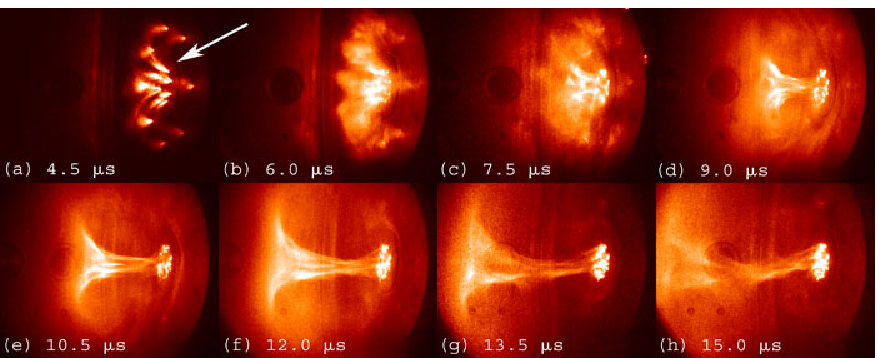}\\
Figure 5
\label{1210}%
\end{figure}

\begin{figure}[ptb]
\includegraphics[width=3truein]{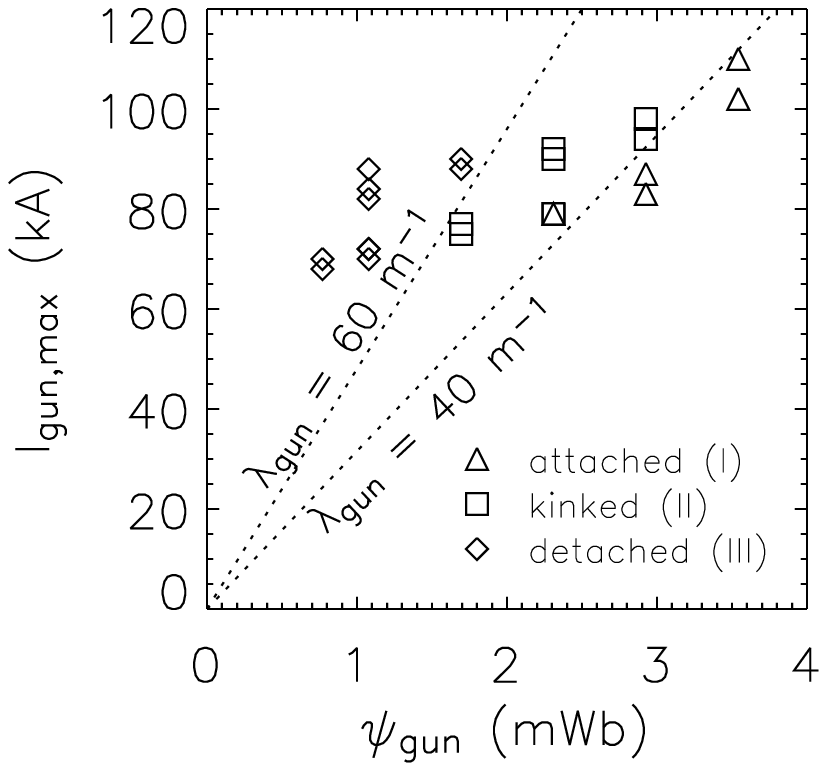}\\
Figure 6
\label{regimes}%
\end{figure}

\begin{figure}[ptb]
\includegraphics[width=3truein]{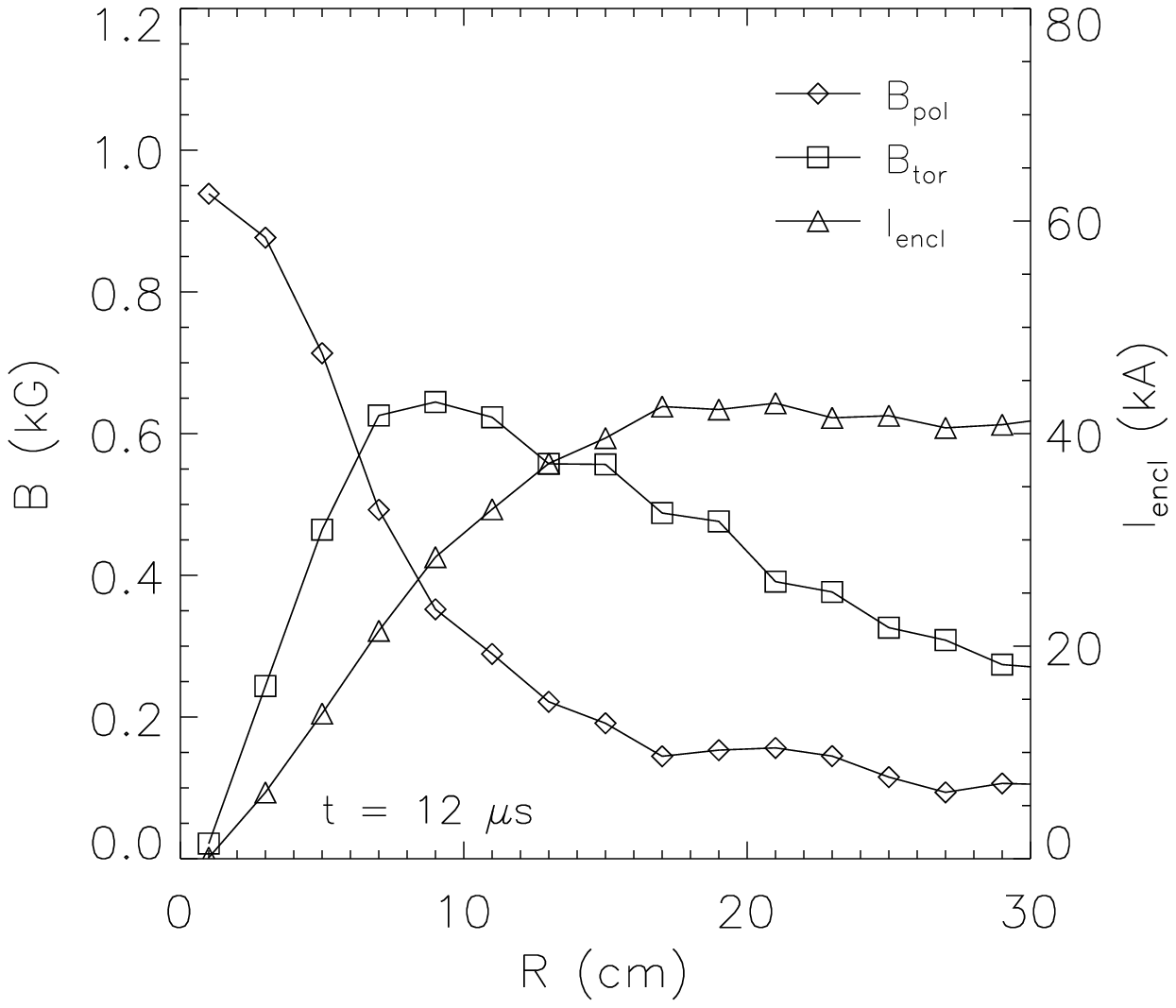}\\
Figure 7
\label{2465_b}%
\end{figure}

\begin{figure}[ptb]
\includegraphics[width=3truein]{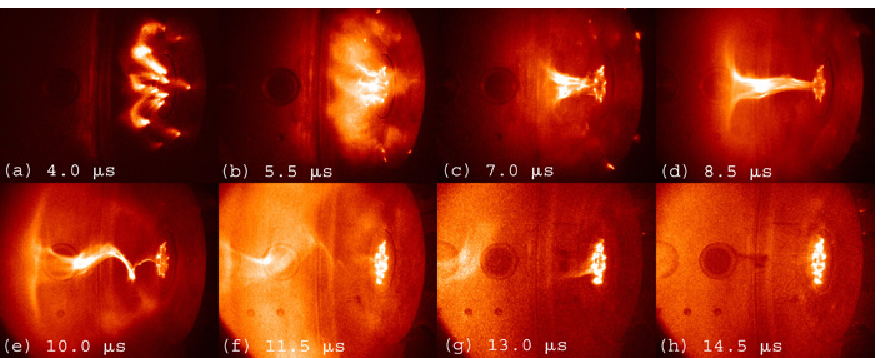}\\
Figure 8
\label{1233}%
\end{figure}

\begin{figure}[ptb]
\includegraphics[width=3truein]{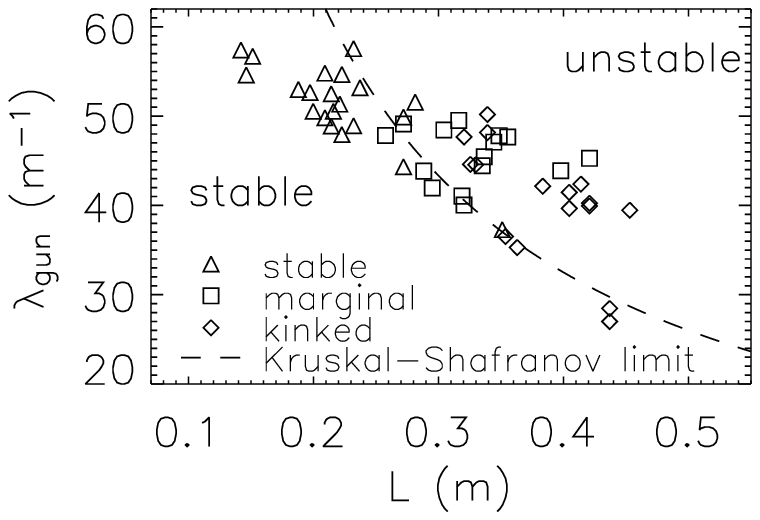}\\
Figure 9
\label{lgun-vs-L}%
\end{figure}

\begin{figure}[ptb]
\includegraphics[width=3truein]{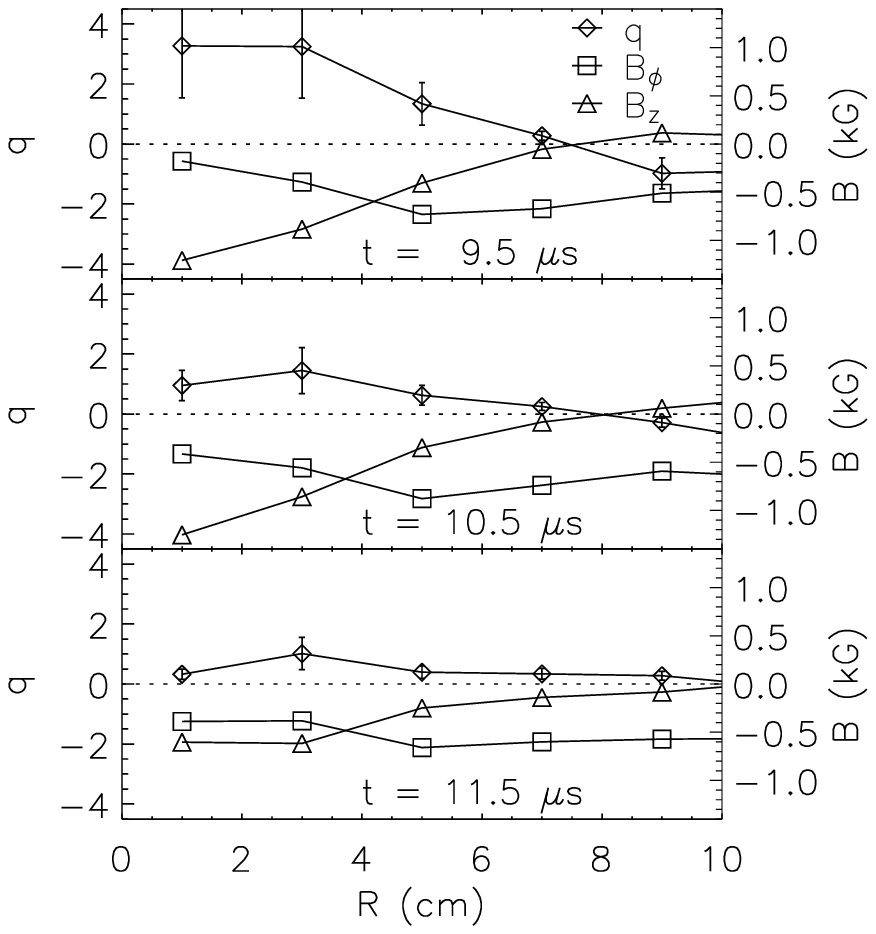}\\
Figure 10
\label{q-profiles}%
\end{figure}

\begin{figure}[ptb]
\includegraphics[width=3truein]{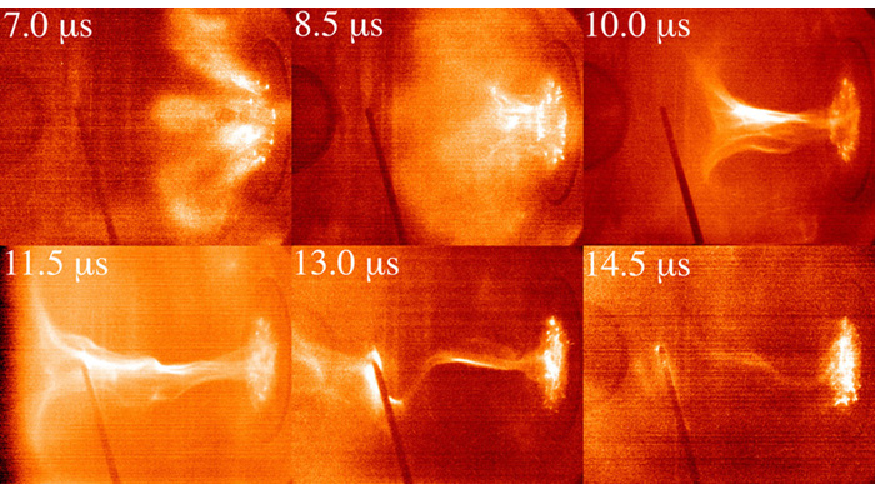}\\
Figure 11
\label{2472}%
\end{figure}

\begin{figure}[ptb]
\includegraphics[width=3truein]{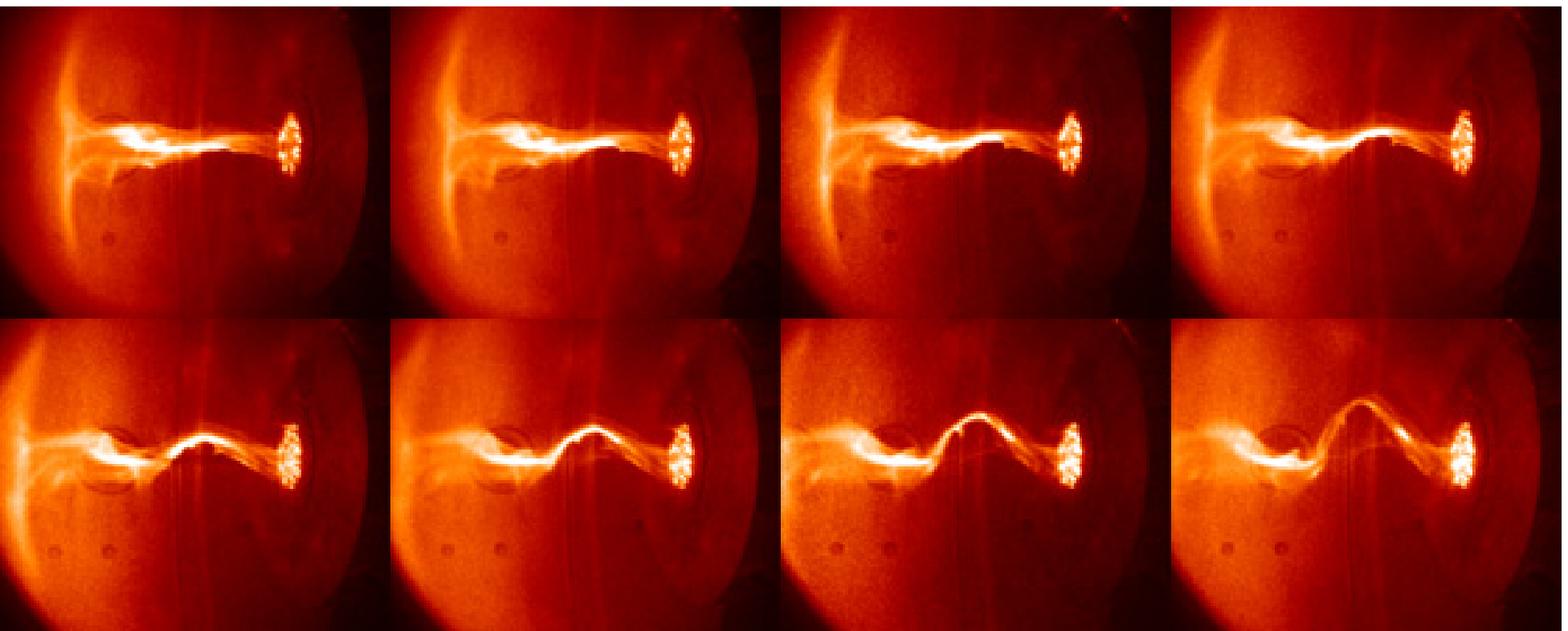}\\
Figure 12
\label{1247}%
\end{figure}

\begin{figure}[ptb]
\includegraphics[width=3truein]{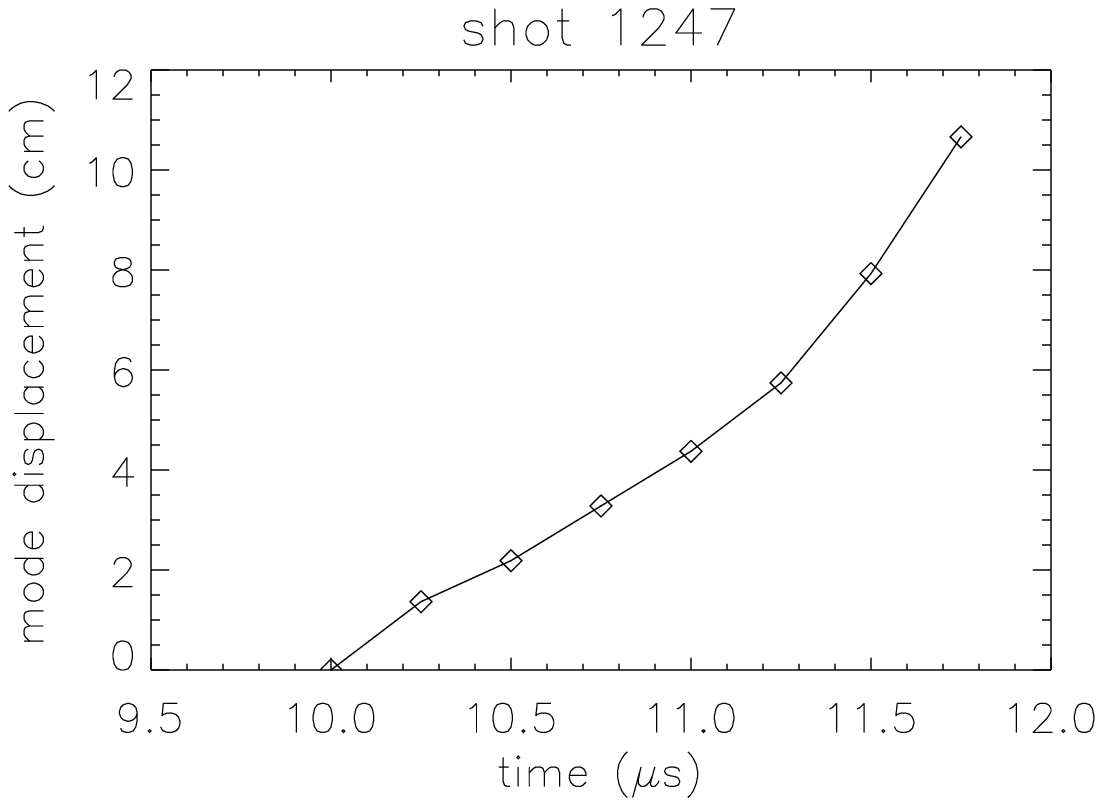}\\
Figure 13
\label{1247-growth}%
\end{figure}

\begin{figure}[ptb]
\includegraphics[width=3truein]{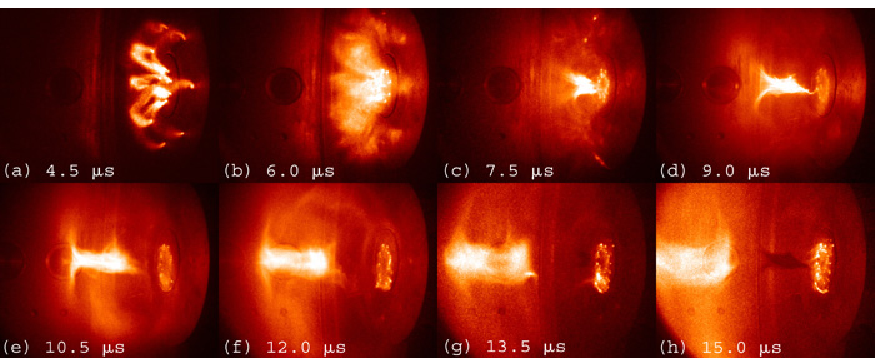}\\
Figure 14
\label{1181}%
\end{figure}

\begin{figure}[ptb]
\includegraphics[width=3truein]{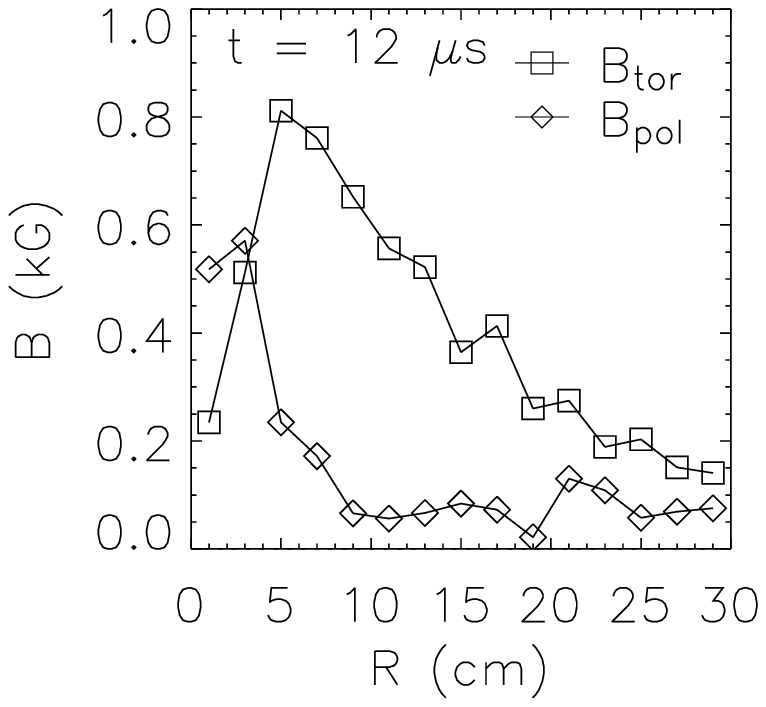}\\
Figure 15
\label{2457}%
\end{figure}

\begin{figure}[ptb]
\includegraphics[width=3truein]{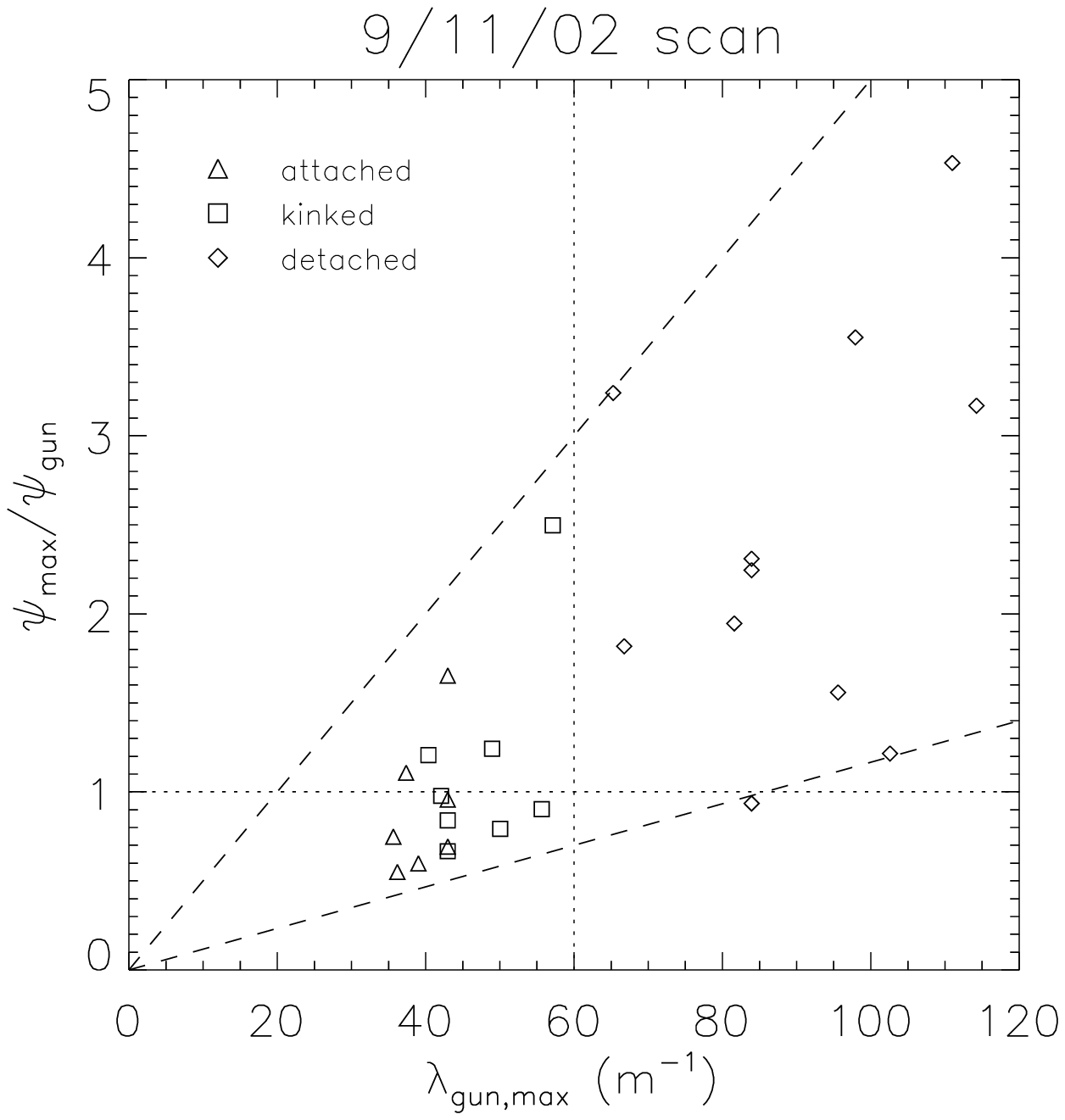}\\
Figure 16
\label{psi-amp}%
\end{figure}

\begin{figure}[ptb]
\includegraphics[width=3truein]{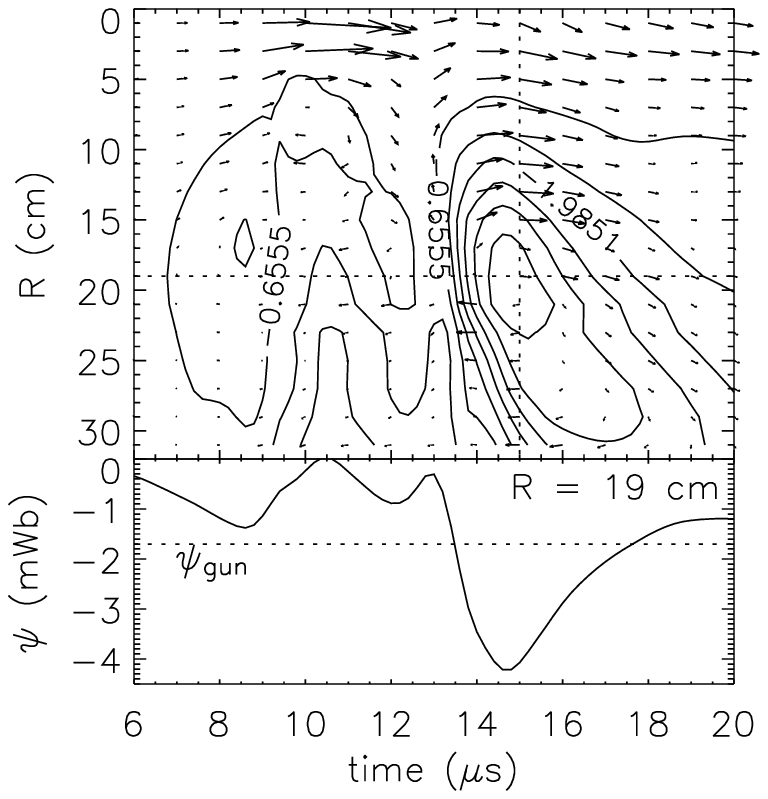}\\
Figure 17
\label{2472_b}%
\end{figure}

\begin{figure}[ptb]
\includegraphics[width=3truein]{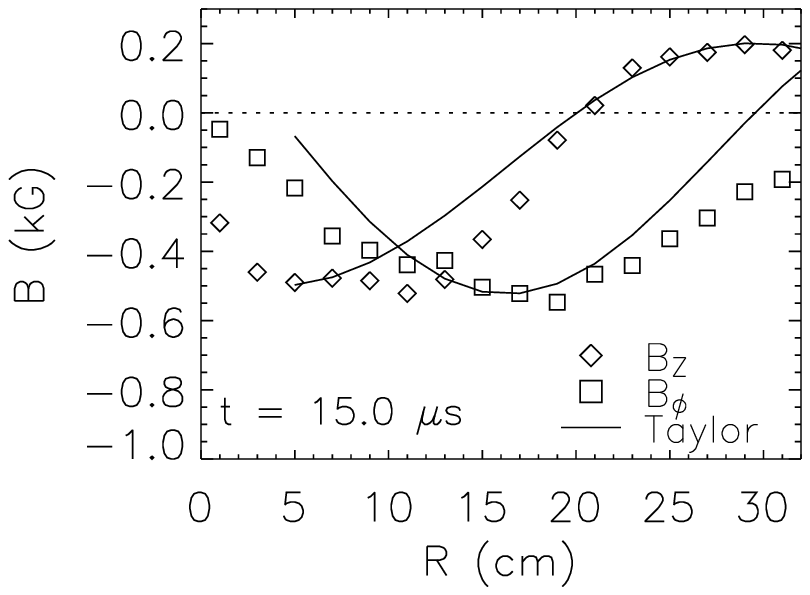}\\
Figure 18
\label{2472_taylor}%
\end{figure}

\end{document}